\documentclass{JHEP3} 
\title{Chiral bosonization for non-commutative fields }
\author{Ashok Das\\ 
Department of Physics and Astronomy, University of
  Rochester, Rochester, NY 14627-0171, USA\\ 
E-mail: \email{das@pas.rochester.edu}}
\author{Jorge Gamboa \\ 
Departamento de F\'{\i}sica, Universidad de Santiago de Chile\\ 
Casilla 307, Santiago 2, Chile\\
E-mail: \email{jgamboa@lauca.usach.cl}}
\author{Fernando M\'endez\\
 INFN, Laboratorio Nazionali del Gran Sasso, SS, 17bis,
  67010 Asergi (L'Aquila),  Italy\\
  E-mail: \email{ fernando.mendez@lngs.infn.it}}
 \author{Justo  L\'opez-Sarri\'on \\
Departamento de F\'{\i}sica Te\'orica, Universidad de Zaragoza,
Zaragoza 50009, Spain\\ 
E-mail: \email{justo@dftuz.unizar.es}}

\abstract{ A model of chiral bosons on a non-commutative  field space is
 constructed and new generalized 
 bosonization (fermionization)  rules for these fields are given. The
 conformal structure of the theory is characterized by a level of the
 Kac-Moody  algebra equal to $(1+ \theta^2 )$ where $\theta$ is the
 non-commutativity 
 parameter and chiral bosons living in a non-commutative
 fields space are described by a rational conformal field theory with
 the central charge of the Virasoro algebra equal to 1.  The
 non-commutative chiral bosons are shown to correspond to a  
 free fermion moving with a speed equal to $ c^{\prime} = c\,
 \sqrt{1+\theta^2}  $ where $c$ is the speed of light.  Lorentz
 invariance remains intact  if $c$  is rescaled by $c \to c^{\prime}$.  The
 dispersion relation for bosons and fermions, in this case, is given
 by $\omega =  c^{\prime} | k|$.}
 
\begin{document}

\section{Introduction}\label{sec:int}
Bosonization is an important theoretical technique in many fields such
as string 
theory \cite{witten}, condensed matter physics  \cite{mattis} and 
mathematical physics \cite{di fran} (for detailed references on
bosonization see \cite{stone}). 
From a purely physical point of view, bosonization is a duality
relation between the strong and the weak coupling limits and, therefore,
it can provide information about the nonperturbative sector of quantum
field theories. 

It  seems quite natural to consider the present relativistic quantum
field theories as 
effective descriptions  where for energies over $10 $ TeV and more,
Lorentz invariance should 
be replaced by another one -up to now unknown symmetry \cite{To,ame,ame1}.
Several  recent events 
violating the GZK bound  of ultra-high energy cosmic rays support this
possibility \cite{swain} and 
a kinematical analysis suggests that for momenta over $10^{16}$ eV, new
non-commutative effects 
could take place and, as a consequence, a tiny violation of the
micro-causality principle would be 
possible \cite{ccgm}. The non-commutative theories, of course, present
examples where Lorentz invariance is violated.
In a different context, Coleman and Glashow  \cite{cole} have also
discussed about 
such a possibility in particle physics processes. However, in the
absence of Lorentz invariance, it is not clear {\it a priori} how to
determine  the correct dispersion relations for bosons
and fermions \footnote{ For a systematic discussion about possible
  violations of Lorentz invariance see  \cite{kost}.}. 

In this sense, an analysis in a $1+1$ dimensional system can be
useful in analyzing this problem. The 
goal of the present paper is to analyze the Abelian
bosonization for a two dimensional 
system and to show how the bosonization formulae are modified in the
non-commutative space 
of fields. More precisely, we will show below that for chiral bosons 
in a non-commutative field space conformal invariance continues to hold 
and that the non-commutativity in the field space leads to \lq free\rq ~
fermions  when chiral bosons are fermionized. The relativistic
symmetry remains intact provided the speed of light is rescaled by
a factor depending on the parameter of non-commutativity. This last
fact can be seen as a transition between the commutative and the 
non-commutative limits.  

The paper is organized as follows. In section II we review very
briefly the analysis of 
chiral bosons  in the commutative field space.  In section III we
deform the canonical Dirac algebra (commutation relations) and analyze
some of the basic features of such a system. In section IV we
generalize the formulae for bosonization (fermionization) for such a
theory. The relativistic invariance of such a theory is discussed
in section V and we conclude with a brief summary in section VI. In
appendix A, we canonically quantize this theory and derive the
time ordered Green's function for such a theory. In appendix B, we
discuss briefly the modified Lorentz transformations in this theory.    

\section{Chiral Bosons in commutative space of fields}\label{int2}

In this section, we will very briefly recapitulate the essential
features of quantization of a chiral boson in the commutative field space
before going into chiral bosons in a non-commutative field space.
Let us consider a massless scalar field in two dimensional Minkowski
space-time. In such a case, the field $\phi$ can be decomposed  as 
\begin{equation}
\phi = \phi_L + \phi_R, \label{1}
\end{equation}
where the $\phi_L(\phi_R)$ describes a left (right)
moving component. The left and right moving fields are also 
represented as $\phi_{\pm}$ (denoting the dependence on the coordinates
$x\pm ct$) which we will use in our discussions.  These individual
components of the scalar field are known as chiral bosons.

In the local description, the dynamics of left and right handed chiral
bosons can be obtained from the action  
\begin{equation}
S_{a} = \int d^2x \left[ a \phi_{a}^{\prime} \dot {\phi}_{a} -
(\phi_{a}^{\prime})^{2}\right], \label{2}
\end{equation}
where the signs $a=\pm$ represent respectively the left and the right
movers \cite{jackiw} and there is no summation over $a$. Here the
prime and the dot denote respectively a derivative with respect to $x$
and $t$.

The action in (\ref{2}) is constrained and the Hamiltonian analysis
leads to the second class constraint
\begin{equation}
\chi_a=  \pi_a - a \phi_a^{\prime},\quad a=\pm, \label{3}
\end{equation}
whose Poisson bracket (with the usual canonical relations) yields
\begin{equation}
\left[ \chi_a (x) , \chi_b (y) \right] = -2a \delta_{ab} \delta^{\prime}
(x-y).  \label{4} 
\end{equation}
With (\ref{3}-\ref{4}), the Dirac brackets for the field variables can
be easily calculated and they take the forms
\begin{eqnarray}
\left[\phi_a (x), \phi_b (y)\right]_D &=&- \frac{a}{2}
\delta_{ab}\epsilon (x- y), \label{5} 
\\
\left[ \phi_a (x), \pi_b  (y)\right] _D&=& \frac{1}{2}\delta_{ab}
\delta (x-y), \label{6} 
\\
\left[ \pi_a(x), \pi_b (y)\right] _D&=& \frac{a}{2}\delta_{ab}
\delta^{\prime} (x-y),\quad a,b=\pm, \label{7} 
\end{eqnarray}
where $\epsilon (x-y)$ represents the alternating step function.
We note from (\ref{5})-(\ref{7}) that the left and the right moving
components 
define independent degrees of freedom and have decoupled Dirac
brackets. 

As is clear from (\ref{2}), the left and the right movers are governed
respectively by the Hamiltonians
\begin{equation}
H_{a} = \int dx\,  (\phi_{a}^{\prime})^{2},\quad a=\pm ({\rm no\,
summation}), \label{8}
\end{equation}
and the Hamiltonian equations of motion take the forms
\begin{eqnarray}
{\dot \phi}_a&=& a \phi^{\prime}_{a}, \nonumber
\\
{\dot \pi}_a &=& a \phi^{\prime}_a, \label{9}
\end{eqnarray}
or equivalently, the second order equation (which can also be seen
from the action in (\ref{2}))
\begin{equation}
\dot{\phi}^{\prime}_a = a \phi^{''}_a.  \label{10}
\end{equation}

The canonical quantization of this system can now be carried out in a
straightforward manner by replacing the Dirac brackets by commutators. 
We will not go into the details of this analysis which has been
discussed  extensively
in the literature, see {\it e.g.}  \cite{review}.

\section{Deforming the Commutators}\label{sec:int3}

We will next study the chiral boson in a non-commutative field
space \footnote{A related problem to our research is, of course, string theory on a $B$-field background see \cite{sei}. This last problem is very similar to the Landau problem in quantum mechanics.}. To that end, we deform the Dirac brackets of (\ref{5}-{7}) and
define the basic commutators (for chiral bosons in a non-commutative
field space) 
\begin{equation}
\left[\phi_a (x), \phi_b (y) \right] = i \Delta_{ab}\epsilon (x-y), \label{11}
\end{equation}
where
\begin{equation}
\Delta_{ab} = \frac{a}{2}\left[\epsilon_{ab} \theta -
  \delta_{ab}\right], \label{12} 
\end{equation}
and $\theta$ is the non-commutative parameter in the field space. The
commutator in (\ref{11}) is manifestly anti-symmetric and satisfies
Jacobi identity. Furthermore, in the limit $\theta=0$, it reduces to
the earlier commutation relations.
We note that unlike (\ref{7}), the left and right moving fields now do
not commute and they become coupled. Indeed, as  
(\ref{11}) and (\ref{12})  mix left and right movers it is quite
natural to assume that the dynamics is described by the following
Hamiltonian operator 
\begin{equation}
H = \sum_a \int dx :(\phi_a^{\prime})^{2}:, \label{ham}
\end{equation}
with appropriate normal ordering.
The operator (\ref{ham}) describes free left and right
movers for the theory defined in a commutative field space. In the
present case, however, it describes an interacting theory because of
the deformed commutation relations \footnote{In the reference \cite{990} was shown --for string theory on the $B$-background-- that the open strings satisfy a mixed boundary condition, which basically mixes the left and right movers. In this sense our results are consistent with \cite{990}.}.

This last feature is very similar to the Landau problem in quantum
mechanics, where the interacting theory is described by a \lq free\rq
~  Hamiltonian
\[
H= \frac{1}{2}p^2,
\]
with the magnetic interaction  included in the commutator, {\it i.e.}
\[
\left[ p_i,p_j\right] = i \epsilon_{ij} B,
\]
where $B$ is the constant magnetic field.

The Heisenberg equations following from (\ref{ham}) are
\begin{equation}
{\dot \phi}_a = 2 \sum_{b=\pm} \Delta_{ab} \phi^{\prime}_b.  \label{13}
\end{equation}
In light-cone coordinates, these equations take the explicit forms
\begin{eqnarray}
\partial_-  \phi_+ &=&  - \theta \phi^{\prime}_- \nonumber
\\
\partial_+  \phi_-&=&  - \theta \phi^{\prime}_+ .\label{13'}
\end{eqnarray}
Equivalently, they can be written as
\begin{eqnarray}
\Box \phi_+&=& \theta^2 \partial^2_x \phi_+ \nonumber
\\
\Box \phi_-&=& \theta^2 \partial^2_x \phi_-, \label{14}
\end{eqnarray}
and, therefore, the effect of the non-commutativity in the field
space appears to modify the speed of light
\begin{equation}
c \rightarrow c^{\prime} =  c\sqrt{1+\theta^2}. \label{15}
\end{equation}
Correspondingly,  the dispersion relation, in this case, becomes
\begin{equation}
E = \pm c\,\sqrt{1+\theta^2} |k| = \pm c^{\prime} |k|. \label{16}
\end{equation}

We note that the matrix $\Delta_{ab}$ in (\ref{12}) is invertible with
an inverse given by
\begin{equation}
\Delta^{-1}_{ab} = \frac{4}{1+\theta^{2}}\,\Delta_{ab}.
\end{equation}
Consequently, the action for the system can also be written as
\begin{equation}
S = \int
d^{2}x\,\left(-\frac{2}{1+\theta^{2}}\,\phi^{\prime}_{a}\Delta_{ab}
\dot{\phi}_{b} -(\phi_{a}^{\prime})^{2}\right),\label{16'}
\end{equation}
where summation over repeated indices is understood. The commutation
relations in (\ref{11}) can be easily checked to follow from this
action. Furthermore, the dynamical equations for the chiral bosons (in
non-commutative field space) now follow to be (which is consistent
with (\ref{13})) 
 \begin{eqnarray}
  \partial_{x}\left(\begin{array}{cc}
  \partial_t-\partial_x&\theta\partial_x \\
  \theta\partial_x&\partial_t+\partial_x
  \end{array}\right)
  \left(\begin{array}{c}
  \phi_+\\
  \phi_-
  \end{array}\right)&=&0.\label{sistema1}
  \end{eqnarray}
The Green's function $G\left(x- y\right)$ can now be defined to
satisfy 
\begin{equation}
\partial_x  \left(\begin{array}{cc}
  \partial_t-\partial_x&\theta\partial_x \\
  \theta\partial_x&\partial_t+\partial_x
  \end{array}\right)~G\left(x- y \right) = \delta^{2} (x-y),  \label{17}
  \end{equation}
with $G\left(x-y\right)$ a $2\times 2$ matrix.
The Green's function is easily determined in the momentum space,
namely, defining
  \begin{equation}
  G (x)=\frac{1}{(2\pi)^2}\int d^{2}k\,\tilde{G}(k)e^{ik\cdot x},
  \label{18}  
\end{equation}
we obtain
\begin{equation}
\tilde{G} (k)=\frac{1}{k\left[(k^{0})^2-(1+\theta^2)k^2\right]}
\left(\begin{array}{cc} 
k^{0}-k&\theta~k\\
\theta~k& k^{0}+k \end{array}\right).
\label{greennc}
\end{equation}

The momentum integration can be carried out in a straightforward
manner using the Feynman prescription. As we will show in appendix A,
the time ordered Green's function of the theory ${\cal G}$ is related
to $G$ (with Feynman's prescription) through the relation
\begin{equation}
{\cal G} = i G \Delta (\theta),\label{relation}
\end{equation}
where $\Delta (\theta)$ has the matrix form
\begin{equation}
\Delta (\theta) = \frac{1}{2}\left(\begin{array}{rr}
-1 & \theta\\
\theta & 1
\end{array}\right),\label{Delta}
\end{equation}
which can also be read out from the defining relation in (\ref{12})
(the matrix is labelled by the indices $\pm$). In the coordinate
space, the time ordered Green's function (as we will show in appendix
A or as can be evaluated directly from (\ref{18}), (\ref{greennc})
and (\ref{relation})) takes the form
\begin{eqnarray}
{\cal G}_{\pm \pm}(x) &=&\mp\frac{1}{8\pi}\left[\left(1\mp
\frac{c^{\prime}}{c}\right) 
  \ln|x-c^{\prime}t|\right.\nonumber\\
 &  & \quad - \left.\left(1\pm \frac{c^{\prime}}{c}\right)
  \ln|x+c^{\prime}t|\right],\nonumber\\
{\cal G}_{+-}(x)&=&\frac{\theta}{8\pi} \ln
\left|\frac{(x-c^{\prime}t)}{(x+c^{\prime}t)}\right|,
\label{21}
\end{eqnarray}
where we have identified
\begin{equation}
c^{\prime}= c\,\sqrt{1+\theta^2},  \label{23}
\end{equation}
with $c$ representing the speed of light and we have omitted terms
involving the massive regulator. 

These last relations provide a complete solution for chiral bosons
living on non-commutative field  
space. In particular, the off-diagonal Green's functions in (\ref{21})
reflect  the effects due non-commutativity in the field space. We note
that when $\theta = 0$, the Green's functions reduce to those of
conventional chiral bosons where the left and the right movers are
decoupled.  

\section{Fermionization of Chiral Bosons in a Non-commutative Field Space }\label{sec:int4}

The fermionization of chiral bosons in a non-commutative field
space is, in principle, straightforward \cite{gh}. In fact, let us define 
\begin{eqnarray}
\psi_+ &=& C~ : e^{i \alpha (\phi_+ -\theta \phi_-)}:, \label{24}
\\
\psi_- &=& C~ : e^{-i \alpha ( \phi_- +\theta \phi_+)}:, \label{25}
\end{eqnarray}
where $C$ is a normalization constant (possibly infinite) and $::$
denotes normal ordering. 
Using (\ref{24}) and (\ref{25}) we can compute the current algebra
following the standard formulae of conformal field theory \cite{di
  fran,spector}. Indeed, using  
\[
:e^{A}::e^{B}: = e^{{<AB>}}:e^{A} e^{B}:, 
 \]
where $A,B$ denote two arbitrary operators, we can determine the
fermionic currents  to be of the forms
\begin{eqnarray}
\psi^{\dagger}_+ (x)\psi_+ (y) &\sim& e^{\alpha^2 \left( 1+ \theta^2
  \right){\cal G}_{++}(x-y)} : e^{i \alpha (\Delta \phi_+  - \theta
  \Delta \phi_- +\cdots)}:,  \nonumber\\ 
\psi^{\dagger}_{+}(x)\psi_{-}(y) & \sim&
  e^{\alpha^{2}\left(1+\theta^{2}\right) {\cal G}_{+-}(x-y)}
  \nonumber \\ 
  &\times& :e^{-\alpha(\phi_+ -\theta \phi_-)(x)}\,e^{-\alpha (\phi_-
  +\theta \phi_+)(y)}:,\nonumber \\
\psi^{\dagger}_{-}(x)\psi_{+}(y) &\sim&
  e^{\alpha^{2}\left(1+\theta^{2}\right) {\cal G}_{-+}(x-y)} 
  \nonumber \\
&\times& :e^{-\alpha(\phi_- -\theta \phi_+)(x)}\,e^{-\alpha(\phi_+
  +\theta \phi_-)(y)}:,\nonumber\\
\psi^{\dagger}_{-}(x)\psi_{-}(y) &\sim&
  e^{\alpha^{2}\left(1+\theta^{2}\right){\cal G}_{--}(x-y)}\,:e^{i
  \alpha (\Delta \phi_-  - \theta  \Delta \phi_+ +\cdots)}:.\nonumber 
  \\
  \label{currents}  
\end{eqnarray}
For convenience, the coefficient $\alpha$ can be chosen to be 
\begin{equation}
\alpha = \frac{4 \pi c^{3}}{(c^{\prime})^3}= \frac{4
  \pi}{(1+\theta^2)^{3/2}},  \label{27}
\end{equation}
although this is not necessary.

We note from (\ref{21}) that, in the limit $x\rightarrow y$, the
Green's functions behave as $\ln |x-y|$ and the
first and the last relations in (\ref{currents}) give (with the use of
the forms  of the Green's function in (\ref{21}))
\begin{equation} 
\psi^{\dagger}_{\pm} (x)\psi_{\pm} (y) =\alpha \left[ \partial_x \phi_{\pm}  -
  \theta  \partial_x \phi_{\mp} + \cdots\right],  \label{26} 
\end{equation}
where  $\cdots$ denote terms (possibly non-regular) that are not relevant for
the calculation of the current algebra. However, we note that
if we take off-diagonal combinations 
such as $\psi^{\dagger}_- \psi_+$, the coefficient in the exponent
is positive and, therefore, in  the limit  $x\rightarrow y$,  
\begin{equation}
\psi^{\dagger}_- (x)\psi_+ (y) =0= \psi^{\dagger}_+ (x)\psi_- (y).  \label{30}
\end{equation}
Thus, we find that the bosonization formulae corrected by
non-commutativity in the $\pm$ sector \cite{mande} take the forms
\begin{equation}
J_\pm =  \frac{\alpha}{\sqrt{\pi}} \left[ \pm\partial_\pm
  \phi_\pm+\theta  \partial_\mp \phi_\mp \right]. \label{29} 
\end{equation}
In such a case, therefore, we cannot have a mass term present in the
Hamiltonian
and the model considered here including the deformed algebra
(\ref{11}) defines a conformal field theory.  

The current algebra can now be calculated easily using the commutation
relations in (\ref{11}) and takes the form 
\begin{eqnarray}
\left[ J_\pm (x), J_\pm (y) \right]  &=&  \pm \frac{i}{2 } k_1~
\delta^{\prime} (x-y), \nonumber\\
\left[ J_+ (x), J_- (y) \right]  &=&  \frac{i}{2 } k_2 \delta^{\prime}
(x-y), \label{31} 
\end{eqnarray}
where the  levels in the algebra in (\ref{31}) are given by
\[
k_1= 1, \,\,\,\,
\,\,\,\,k_2=  \theta 
\]
Equation  (\ref{31}) represents two coupled $U(1)$ 
Kac-Moody algebras where the levels are different for the mixed and
the unmixed 
commutators.  However, this is not a problem since the algebra can be
diagonalized by going to a different basis.  

In order to do that, we  note that  although each
coefficient in the fermionic current is different, one can
posit the general Hamiltonian operator  
\begin{equation}
H = \int dx: \left[  J^2_+ + J^2_- + \beta J_+ J_-\right]:,\label{32}
\end{equation}
where $\beta$ is a constant. However, we note that the current algebra
is  diagonalized for $\beta =0$. Therefore, we will take this value
for $\beta$. We observe that  if  instead
of the currents $J_\pm$ we consider the new set of currents
$j_\pm$  defined as
\begin{eqnarray}
j_+ &=& J_+\cosh \omega  - J_- \sinh \omega, \nonumber
\\
j_-&=& J_+\sinh \omega  + J_- \cosh \omega,  \label{33}
\end{eqnarray}
then, in this new basis, the current algebra has the form 
\begin{eqnarray}
\left[j_{\pm} (x) , j_{\pm} (y) \right] &=& \pm
\frac{i}{2}(1+\theta^2)~ \delta^{\prime} (x-y)\label{34} 
\\
\left[j_- (x), j_+ (y)\right] &=& 0,  \label{35}
\end{eqnarray}
provided (this follows from (\ref{35})) 
\[
\sinh 2 \omega  +\theta=0.
\]

It follows now from (\ref{34})-(\ref{35}) that the  new levels  of the
Kac-Moody algebra are 
\begin{equation}
k_1=1 + \theta^2, \,\,\,\,\,\,\,\,\,\,\,\, k_2 =0, \label{e1f}
\end{equation}
and that the model of chiral bosons in a non-commutative field space
is a conformal field theory.
The Hamiltonian in terms of these new currents is 
\begin{equation}
H_j =\frac{1}{\sqrt{1 + \theta^2}} \int dx :\left[ j^2_+ + j^2_-
  \right]:.  \label{new1} 
\end{equation} 
Using the bosonization rules, we can write the Hamiltonian in terms
of fermionic fields as
\begin{equation}
H_F = \frac{1}{\sqrt{1 + \theta^2}}\int dx :\left[i\psi^{\dagger}_-
  \frac{d}{dx}  \psi_- - i\psi^{\dagger}_+  
\frac{d}{dx}  \psi_+\right]:.   \label{new2}
\end{equation} 

The fermionic two-point functions for this system can be computed
straightforwardly  using the bosonization rules (\ref{24})-(\ref{25})
and (\ref{29}). Indeed, we have  
\begin{equation}
<\psi^{\dagger}_\pm (x) \psi_\pm (0)>  \sim
\frac{1}{(x+vt)^{\frac{\gamma_\pm}{2} } 
(x-vt)^{\frac{\gamma_\mp}{2}}}, \label{42}
\end{equation}
where
\begin{equation}
\gamma_{\pm} = 1 \pm \frac{1}{\sqrt{1+\theta^{2}}} = 1 \pm
\frac{c}{c^{\prime}}, 
\end{equation}
and in the limit  $\theta \to 0$, the conventional propagator is
correctly recovered. The remaining Green's functions have the forms 
\begin{equation}
<\psi^{\dagger}_\pm (x) \psi_\mp (0)> \sim  \left(\frac{x+vt}{x-vt}
\right)^{\frac{\theta}{2(1+\theta^2)}}, \label{555} 
\end{equation}
which becomes trivial in the commutative limit. 

From (\ref{42}), we see that although (\ref{new2}) formally
describes a \lq free\rq  ~fermion, the Green's function  (\ref{42})
reflects an interaction due to non-commutativity in the field space that
disappear in the limit $\theta \to 0$. However, even in the
non-commutative regime Lorentz invariance remains intact  if we scale
the speed of light $c$ to $c^{\prime}$ as defined in (\ref{15}) and as we
discuss   in the next section. In this sense, the system behaves like
a dielectric with a dielectric constant related to the parameter of
non-commutativity.   

Finally we would like to close this section with some comments about
the nature of the conformal field theory and the Virasoro algebra
associated with this system.  
Given the Kac-Moody algebra in (\ref{34})-(\ref{35}), we can
compute the Virasoro algebra in a straightforward manner. Let us define   
\begin{equation}
L_n = \frac{1}{\sqrt{1+\theta^2}} \int_{-\pi}^\pi dx\, e^{inx}
:j^2(x):. \label{4111}  
\end{equation}
It can now be easily seen that
\begin{equation}
\left[ L_n, L_m\right] = (n-m)L_{n+m} + \frac{{\tilde c}}{12} n(n^2
+1) \delta_{n+m}, \label{4222} 
\end{equation}
where 
\begin{equation}
{\tilde c}= 1, \label{4333}
\end{equation}
is the central charge of the Virasoro algebra \cite{callan}.  Thus, we conclude
that model of chiral bosons in a non-commutative field space defines a
rational conformal field theory with the Kac-Moody level
$(1+\theta^{2})$ and the central charge of the Virasoro algebra 1.  

\section{Relativistic symmetries and Lorentz Invariance}\label{int5}

In the last section we found that chiral bosons in the non-commutative
field space can be mapped to a \lq free \rq ~fermion with Green's
functions given by (\ref{42})-(\ref{555}). This result is surprising
and unexpected. In addition,
the commutator (\ref{11}) explicitly breaks conventional  
Lorentz invariance and, therefore, the main question is whether there
are other generalized Lorentz symmetries in the non-commutative regime.  

In appendix B, we discuss how Lorentz transformations are
modified if the speed of light is changed as in (\ref{15}).   
We note here that theories in two-dimensions are very special. Indeed,
the parameter
$\theta$ is dimensionless and, therefore, the only effect produced by
non-commutativity is to modify the speed of light.  However, this
modification is non-trivial because although the only change in Lorentz
transformations due to this is to make the replacement   
(see eq. (\ref{15})) 
\begin{equation}
c \rightarrow c^{\prime}= c\,\sqrt{1+\theta^2}, \label{rede} 
\end{equation}
the physical interpretation is non-conventional. 
Namely, the rescaling in $c$ leads us to define the ratio
\begin{equation}
n = \frac{c}{c^{\prime}} = \frac{1}{\sqrt{1+\theta^2}}, \label{300}
\end{equation}
which can be thought of as a two-dimensional index of refraction which
is directly related to the parameter of non-commutativity in the field
space.  Thus, we can think of a sort of transition between the
commutative and the  non-commutative regimes
where the velocity of the photon, for example,  increase when it leaves
the commutative regime.    

This interpretation, however, cannot be extended to higher dimensions
because, in such case,  the dimension of $\theta$ is nontrivial in
units of energy. Consequently, a new non-commutative dimensional
parameter with inverse dimensions needs to be introduced for such a
phenomenon. As has been discussed in
previous papers \cite{ccgm}, one can add a new non-commutative
parameter in the commutators of momenta in the phase space.  If we
denote such a  parameter as $B$ (in analogy to motion in a magnetic
field), then $\theta$ and $B$ play the roles of
the ultraviolet and infrared cutoffs as discussed in
 \cite{ccgm}. When  
the product  $\theta B$ is dimensionless, the limits $\theta B >>1$
and $ \theta B <<1$ define the ultraviolet and infrared regimes
respectively.  The case $\theta B =1$ is singular and it would
correspond to the critical point suggested above \footnote{This limit
  was discussed in \cite{poly} and later in many references on
  non-commutative quantum mechanics \cite{jg}.}. Thus, in higher
dimensions, the phase space considerations are necessary if we want
to retain extended relativistic invariance. This can, in fact, be
argued in favor of extra dimensions.  
We would also like to emphasize that our calculations have a direct
application to cosmology where some authors have argued that the
speed of light could be bigger than $c$ in the first instants of
the universe \cite{barrow}. 

If the chiral bosons are equivalent to a
massless \lq free\rq ~fermion, then we can imagine the Hamiltonian
(\ref{32}) to come from  
\begin{equation}
{\cal L} = {\bar \psi} i\partial\!\!\!\slash\,
\psi , \label{411} 
\end{equation}
where the temporal component of the derivative in
$\partial\!\!\!\slash$ should be understood as $\partial_0
=\frac{1}{c^{\prime}} \frac{\partial
}{ \partial t}$. 
With this interpretation, the dispersion relations for bosons and
fermions hold automatically and can be computed directly from $1/
\partial\!\!\!\slash$ leading to  
\[
\omega = |E| = c^{\prime} |k|, 
\]
in full agreement with  (\ref{16}).  

Thus, relativistic invariance implies that there is no difference
in the dispersion relation for bosons and fermions and is determined 
in terms of the rescaled speed of light (\ref{rede}). 
For a  non-relativistic system, on the other hand, $c^{\prime}$ can simply
be thought of as the Fermi velocity
with the system discussed above describing a fermion embedded in a
medium.

\section{Conclusions}\label{sec:int5}

In this paper we have discussed the dynamics of chiral bosons defined
in a 
non-commutative field space and have generalized the Abelian
bosonization to such a case. Our results go over smoothly to the
commutative limit and contain the Mandelstam formulae as a special
case.  As has been discussed in previous papers  
 \cite{ccgm}, the modification of the canonical commutation relations
is equivalent to adding interactions to the Lagrangian (or the
Hamiltonian) with the standard commutation relations. This is
easily seen from the  form of the action in (\ref{16'}). A constraint
analysis of this system leads to the commutators in (\ref{11}) as well
as the Hamiltonian equations discussed earlier. In spite of the
non-commutativity in the field space, the model possesses Lorentz
invariance with a scaled speed of light. In this sense, the behavior
of the system is reminiscent of a dielectric. The current algebra of
the system shows that the system is described by a rational
conformal field theory with the Kac-Moody level $(1+\theta^{2})$ and
the central charge in the Virasoro algebra equal to 1. We would like
to point out here that more recently 
chiral bosons in a non-commutative space-time have been discussed in
 \cite{aleman} from a different perspective. However, our approach has
been to analyze the model of chiral bosons with non-commutativity in
the field space and, consequently, our approach is quite distinct.

We would like to thank G. Amelino-Camelia, J. L. Cort\'es, A. Grillo
and M. Loewe  for useful discussions. 
This work was supported in part by US DOE Grant number DE-FG
02-91-ER40685, FONDECYT 1010576 and MECESUP-USA-109. J. L-S. thanks
Ministerio de Educacion y Cultura (grant  AP99 07566588) and
O.N.C.E. for support and  F.M. would thanks  to INFN for a posdoctoral
Fellowship. 

\appendix
\section{Derivation of the Green's Functions}

In this appendix we will derive the forms of the time ordered Green's
function discussed in  (\ref{21}) starting from a canonical
quantization of the theory. Let us start with the definition of the
Feynman Green's function 
\begin{eqnarray}
{\cal G}_{ab}(x,t) &= &
  \theta(t)\langle\Omega\vert\phi_a(t,x)\phi_b(0)\vert\Omega\rangle 
\nonumber \\
& &\quad +
  \theta(-t)\langle\Omega\vert\phi_b(0)\phi_a(t,x)\vert\Omega\rangle, 
\label{fey}
\end{eqnarray}
where $\theta(t)$ is the step function and $\Omega$ denotes the
vacuum of the theory. This defines the time ordered Green's functions
of the theory. 
Using (\ref{sistema1}) as well as (\ref{11}), it is straightforward to
check from the definition in (\ref{fey}) that 
\begin{equation}
{\cal D}{\cal G}(t,x)=i\Delta(\theta)\delta^{2}(x),
\label{diferencial}
\end{equation}
where ${\cal D}$ is the differential operator given in
(\ref{sistema1}), and $\Delta(\theta)$ is defined in (\ref{Delta}).
Comparing with (\ref{17}), we identify (in the matrix notation)
\[
{\cal G} (x) = iG (x) \Delta (\theta),
\]
as pointed out earlier.

To calculate the time ordered Green's function, let us decompose the
chiral boson fields in the non-commutative field space in terms of
plane waves as
\begin{eqnarray}
\phi_+(x) & = & \sqrt{\frac{c^{\prime}}{c}}\int_0^{\infty}\frac{dk}{\sqrt{4\pi
 |k^{0}|}}\left\{a(k)\,e^{-ik\cdot x}+a^{\dag}(k)\,e^{ik\cdot
  x}\right\},\nonumber\\
& & \\ 
\phi_-(x)& = & \sqrt{\frac{c^{\prime}}{c}}\int_0^\infty\frac{dk}{\sqrt{4\pi
 |k^{0}|}}\left\{b^\dag(k)\,e^{-ik\cdot x}+b(k)\,e^{ik\cdot
  x}\right\},\nonumber 
\end{eqnarray}
where $|k^{0}|=c^{\prime}|k|$ and the creation and annihilation
operators $a$, $b$, $a^\dag$ and $b^\dag$ satisfy, 
\begin{eqnarray}
\left[a(k),a^\dag(k^\prime)\right]&=&\left[b(k),b^\dag(k^\prime)\right]=
\delta(k-k^\prime),\\
\left[a(k),b(k^\prime)\right]&=&\left[b^\dag(k),a^\dag(k^\prime)\right]=
-\theta\delta(k-k^\prime),
\end{eqnarray}
with all other commutators vanishing. This guarantees the commutation
relations between the fields in (\ref{11}).
The Hamiltonian of the theory in (\ref{ham}) can now be expressed as,
\begin{equation}
H=c\int_0^\infty dk\,k\left\{a^\dag(k)\,a(k)+b(k)\,b^\dag(k)\right\}
\label{Hamilton}
\end{equation}
However, we note that since the operators $(a,a^{\dagger})$ do not
commute with $(b,b^{\dagger})$, they cannot be thought of as
raising/lowering energy eigenstates. Consequently, we need to
diagonalize the basis.

The diagonalization is fairly straightforward. Defining a new set of
creation and annihilation operators as
\begin{eqnarray}
c^\dag(k)&=&\sqrt{\frac{c}{2c^{\prime}}}\left[{\gamma_-}^{1/2}\,a(k)
+\gamma_+^{1/2}\,b^\dag(k)\right]~\\
d(k)&=&\sqrt{\frac{c}{2c^{\prime}}}\left[{\gamma_+}^{1/2}\,a(k)
-{\gamma_-}^{1/2}\,b^\dag(k)\right],\nonumber  
\\ 
\end{eqnarray}
where as defined earlier,
$$
\gamma_\pm=1\pm\frac{1}{\sqrt{1+\theta^2}} = 1\pm \frac{c}{c^{\prime}},
$$
it is easy to check that they satisfy
\begin{eqnarray}
\left[c(k),c^\dag(k^\prime)\right]&=&\left[d(k),d^\dag(k^\prime)\right]=
\delta(k-k^\prime),\\
\left[c(k),d(k^\prime)\right]&=&\left[c^\dag(k),d^\dag(k)\right]=
\left[c(k),d^\dag(k^\prime)\right]=0. \nonumber 
\end{eqnarray}
In this new basis, the Hamiltonian (\ref{Hamilton}) takes the form
\begin{equation}
H=c^{\prime}\int_0^\infty dk\,k\left\{c(k)\,c^\dag(k)+d^\dag(k)\,d(k)\right\},
\label{Hamilton1}
\end{equation}
with
$$
c^{\prime}=c\,\sqrt{1+\theta^2}.
$$

The field decomposition in this new basis is easily obtained to be
\begin{eqnarray}
\phi_+(t,x)&=& \sqrt{\frac{c^{\prime}}{c}}\int_0^{\infty}\frac{dk}{2\sqrt{2\pi
    k}}
\biggl[\gamma_-\,c^\dag(k)\,e^{-ik(x-c^{\prime}t)} \nonumber 
\\
& & \quad + \gamma_+\,d(k)\,e^{-ik(x+c^{\prime}t)}\biggr]+{\rm c.c.} \\
\phi_-(t,x)&=&\sqrt{\frac{c^{\prime}}{c}}\int_0^{\infty}\frac{dk}{2\sqrt{2\pi
    k}}\biggl[\gamma_+\,c^\dag(k)\,e^{-ik(x-c^{\prime}t)} \nonumber \\
& &\quad - \gamma_-\,d(k)\,e^{-ik(x+c^{\prime}t)}\biggr]+{\rm c.c.}. 
\end{eqnarray}
Furthermore, using the integral representation for the step function,
$$
\theta(t)=\int_{-\infty}^\infty\frac{d\omega}{2\pi i}\frac{e^{i\omega
    t}}
{\omega -i\epsilon},
$$
 it is a matter of algebra to check that the time ordered Green's
 function in (\ref{fey}) is given by,
\begin{eqnarray}
{\cal G}_{++}(t,x)&=&-\frac{c^{\prime}}{8\pi c}\left\{\gamma_+{\rm
  log}\vert x+c^{\prime} t\vert+\gamma_-\,{\rm log}\vert
  x-c^{\prime}t\vert\right\}, \nonumber \\
{\cal G}_{--}(t,x)&=&-\frac{c^{\prime}}{8\pi c}\left\{\gamma_-{\rm
  log}\vert x+c^{\prime} t\vert+\gamma_+\,{\rm log}\vert x-c^{\prime}
  t\vert\right\}, \nonumber \\
{\cal G}_{+-}(t,x)&=&{\cal G}_{-+}(t,x)=-\frac{\theta}{8\pi}{\rm log}
\left\vert\frac{x-c^{\prime} t}{x+c^{\prime} t}\right\vert .
\end{eqnarray}

\section{Modified Lorentz Transformations}
Now we discuss the question of relativistic covariance of this
theory. In order to do that, let us consider the equation of motion
(\ref{13}), {\it i.e.}  
\begin{eqnarray}
\partial_0\phi_+-\partial_1\left(\phi_+-\theta\phi_-\right)=0\\
\partial_0\phi_-+\partial_1\left(\phi_-+\theta\phi_+\right)=0
\end{eqnarray}
where $x^0=ct$ and $x^1=x$. 

Now we want to identify a symmetry transformation of the fields
related to the linear coordinates transformations given by, 
\begin{eqnarray}
x^{0\,\prime}=a x^0+b x^1\\
x^{1\,\prime}=p x^0+q x^1
\end{eqnarray}
where the coefficients $a$, $b$, $p$ and $q$ will be determined below. If
we make a coordinate transformations as above, then the derivatives will
transform as,
\begin{eqnarray}
\partial_0\rightarrow a\partial_0+p\partial_1 \\
\partial_1\rightarrow b\partial_0+q\partial_1
\end{eqnarray}
and hence, the equations of motion will take the forms,
\begin{eqnarray}
\partial_0\left[(a-b)\phi_++b\theta\phi_-\right]+\partial_1\left[(p-q)\phi_++q
\theta\phi_-\right]=0,
\nonumber 
\\
\label{b1}
\\
\partial_0\left[(a+b)\phi_-+b\theta\phi_+\right]+\partial_1\left[(p+q)\phi_-+q
\theta\phi_+\right]=0, \nonumber
\\
\label{b2}
\end{eqnarray}

The equations of motions will be invariant under this 
transformations, if the fields $\phi_{\pm}$ transform suitably. 
Indeed, by comparison we find from (\ref{b1}), (\ref{b2}) and
(\ref{13}) that 
\begin{eqnarray}
\tilde\phi_+=(a-b)\phi_++b\theta\phi_-\\
\tilde\phi_-=(a+b)\phi_-+b\theta\phi_+
\end{eqnarray}
and, therefore, one have
\begin{eqnarray}
\tilde\phi_+-\theta\tilde\phi_-=(q-p)\phi_+-q\theta\phi_-\\
\tilde\phi_-+\theta\tilde\phi_+=(p+q)\phi_-+q\theta\phi_+.
\end{eqnarray}
Multiplying the second equation by $\theta$ and adding to the first
one, we get
$$\tilde\phi_+=\left[q-p\frac{1}{1+\theta^2}\right]\phi_++p\frac{\theta}
{1+\theta^2}\phi_-.$$ 

Similarly, by multiplying the first equation by $\theta$ and subtracting 
the second equation, we obtain
$$\tilde\phi_-=p\frac{\theta}{1+\theta^2}\phi_++\left[q+p\frac{1}
{1+\theta^2}\right]\phi_-.$$
These equations are compatible with the other expressions for
$\tilde\phi_\pm$, if $a=q$ and $b=\frac{p}{1+\theta^2}$.

Finally, if we rescale the unit of time such that,
$$x^0\rightarrow \frac{x^0}{\sqrt{1+\theta^2}}, $$
the coordinate transformation is orthogonal, {\it i.e.} 
\begin{eqnarray}
x^{0\,\prime}=ax^0+\frac{p}{\sqrt{1+\theta^2}}x^1,\\
x^{1\,\prime}=\frac{p}{\sqrt{1+\theta^2}}x^0+ax^1.
\end{eqnarray}
Using the same notation as in the standard Lorentz transformation where
$a\equiv\gamma$ and $p\equiv -\gamma\beta$ with 
$\beta=\frac{v}{c}$ with $c$ the speed of light, it is easy to
check that the invariance of the space-time interval implies that,
$$\gamma =\frac{1}{\sqrt{1+\frac{v^2}{c^2(1+\theta^2)}}},$$
so that the final expression for the symmetry transformations are,
\begin{eqnarray}
x^{0\,\prime}=\frac{x^0-\frac{v}{c^\prime}x^1}{\sqrt{1-\frac{v^2}{c^{\prime
	2}}}},\\ 
x^{1\,\prime}=\frac{x^1-\frac{v}{c^\prime}x^0}{\sqrt{1-\frac{v^2}{c^{\prime
	2}}}},
\end{eqnarray}
where $c^\prime=c\,\sqrt{1+\theta^2}$. 
Namely, the new symmetry transformations are the same as Lorentz
transformations with a modified speed of light.

\end{document}